\documentclass[twocolumn,showpacs,preprintnumbers,amsmath,amssymb]{revtex4}
\usepackage{graphicx, epsfig, psfrag}

\begin{document}   
   
\title{Modified Gravity via Spontaneous Symmetry Breaking} 
\author{B. M. Gripaios}  
\email{b.gripaios1@physics.ox.ac.uk}  
\affiliation{Rudolf Peierls Centre for Theoretical Physics, University of Oxford,
1 Keble Rd., Oxford OX1 3NP, UK}    
\begin{abstract}   
We construct effective field theories in which gravity is modified via spontaneous breaking of local Lorentz invariance. This is a gravitational analogue of the Higgs mechanism. These theories possess additional graviton modes and modified dispersion relations. They are manifestly well-behaved in the UV  and free of discontinuities of the van Dam-Veltman-Zakharov type, ensuring compatibility with standard tests of gravity. They may have important phenomenological effects on large distance scales, offering an alternative to dark energy. For the case in which the symmetry is broken by a vector field with the wrong sign mass term, we identify four massless graviton modes (all with positive-definite norm for a suitable choice of a parameter) and show the absence of the discontinuity. 
\end{abstract}   
\maketitle   

\section{Introduction}
There is, by now, convincing evidence for new physics associated with the gravitational interaction on cosmological distance scales \cite{Perlmutter:1998np}. It is fashionable to attribute these phenomena to esoteric sources of energy-momentum \cite{Straumann:2002he}, but it is conceivable that gravity itself should be modified on large scales by adding terms with two or fewer derivatives to the action for General Relativity \cite{Fierz:1939ix}. However, any such term (bar the cosmological term) breaks local Lorentz invariance, with two generic consequences. Firstly, because the symmetries of the standard and modified gravity theories are not the same, there is no way to pass continuously from one to the other. This discontinuity necessarily implies a contradiction with solar-system tests \cite{vanDam:1970vg}. Secondly, it is symmetry which guarantees that General Relativity (GR) is well-behaved in the ultra-violet and ensures that strong-coupling effects are postponed to the Planck scale ${M_P}^{-1} \sim 10^{-33} \mathrm{cm}$. In modified gravity theories, strong-coupling typically sets in at far lower scales, contradicting table-top tests \cite{Arkani-Hamed:2002sp}.

To circumvent these difficulties, we consider theories in which local Lorentz symmetry is present ab initio, but is spontaneously broken in vacuo, leading to a modified theory of gravity. This is a gravitational analogue of the Higgs mechanism \cite{Abers:1973qs} in gauge theories \cite{Arkani-Hamed:2003uy}. We start from an effective theory (with UV cutoff $M$) for a field transforming in some representation of the Lorentz group with a `wrong sign' mass term. This field acquires a vacuum expectation value (\emph{vev}), breaking Lorentz invariance; the coupling to gravity results in additional two- and zero-derivative terms appearing in the gravitational effective action, suggesting modified dynamics on large scales. Since the symmetries of the alternative gravity theories are the same, no discontinuity arises as the \emph{vev} tends to zero and GR is recovered. Moreover, the UV behaviour is evident: the cut-off is given by $\mathrm{min}(M,M_P)$.  Conflict with experiment is thus easily avoided. 

Because Lorentz invariance is broken in the vacuum, there are marked contrasts with the gauge-theory Higgs mechanism, in which the gauge fields corresponding to broken generators of the gauge group acquire a third polarization and a mass. Here, for example, graviton modes with different polarizations may have different dispersion relations $\omega(\mathbf{k})$, where the 4-momentum is $k^{\mu}=(\omega,\mathbf{k})$, which may themselves be Lorentz-variant. Also, one no longer has a Lorentz-invariant definition of mass as $\sqrt{k^{\mu}k_{\mu}}$.  One can instead define a mass gap as $\omega (\mathbf{k}=\mathbf{0})$, but this is frame-dependent. Finally, although terms without derivatives appear in the gravity action, there is no theorem that graviton modes acquire a mass gap.

In the remainder of this article, we consider the simplest example where a vector field acquires a \emph{vev} $Mn^{\mu}$ and analyze small fluctuations around a flat background. There are four massless graviton modes and corrections to GR are of order $M^2/M_{P}^{2}$. One can choose $ 10^{-31} \mathrm{cm} \lesssim M^{-1} \lesssim 10^{-2} \mathrm{cm}$ without contradicting existing experiments in approximately-flat spacetimes \cite{Will:1993ns,note1,note9}. Moreover, (with a suitable choice of a parameter) all the physical graviton modes have positive-definite norm, as does the Hilbert space of physical states.

\section{Vector Field Breaking}
The effective action for Einstein gravity coupled to a vector boson $A^{\mu}$ can be written, up to a total derivative, as
\begin{widetext}
\begin{multline} \label{act}
S = \int d^4 x \sqrt{-g} \bigg[ M_{Pl}^{2} R 
- \frac{1}{4g^2} F_{\mu\nu}F^{\mu \nu}
+ \frac{\alpha_1}{2} R_{\mu\nu} A^{\mu}A^{\nu} 
+ \frac{{\alpha_2}}{2} (D_{\mu}A^{\mu})^2 
+\frac{\beta_1}{2 M^2} F_{\mu\nu}F^{\mu\sigma} A_{\sigma} A^{\nu}
+\frac{\beta_2}{2 M^2} D_{\mu} A^{\mu} D_{\nu} A^{\sigma} A^{\nu}A_{\sigma} \\
+\frac{\beta_3}{2 M^2} D_{\mu} A^{\nu} D_{\sigma} A_{\nu} A^{\mu}A^{\sigma} 
+\frac{\beta_4}{2 M^4} D_{\mu} A^{\nu} D_{\sigma} A^{\rho} A^{\mu}A^{\sigma}A_{\nu}A_{\rho}
- \frac{\gamma}{2} (A_{\mu}A^{\mu} - M^2 n_{\mu}n^{\mu})^2 
 + \dots \bigg],
\end{multline}
\end{widetext}
where $g \ll 1$ is the vector boson coupling and ${\alpha_i},{\beta_j},\gamma,n^{\mu} n_{\mu}$ are dimensionless coefficients of order unity. The ellipsis denotes terms suppressed by powers of $M$ or $M_{P}$ \cite{note2}. 

\subsection{Without gravity}In the limit $M_P \rightarrow \infty$, gravity is decoupled and we have the spacetime analogue of the Goldstone mechanism \cite{Low:2001bw}. The action (\ref{act}) has extrema at $A^{\mu}=0$ and $A^{\mu} = Mn^{\mu}$. To find the propagating degrees of freedom at tree-level, we expand the effective Lagrangian to quadratic order in fluctuations around the vacuum. Propagating modes correspond to the residues at the poles of the two-point Green's function \cite{'tHooft:1973pz}. Equivalently, one can find the eigenvectors of the kernel of the quadratic Lagrangian with vanishing eigenvalue (henceforth referred to as \emph{null eigenvectors}). 

 If $\gamma n^2 >0 $, the fluctuations about the vacuum $\langle A^{\mu}\rangle =0$ correspond to a vector boson with mass-squared $2\gamma g^2 M^2 n^2$ (the mass is below the cut-off $M$ for $g \ll 1$), with three modes polarized orthogonally to $k^{\mu}$. The fourth mode is polarized along $k^{\mu}$, but its eigenvalue does not vanish for any value of its energy below the cut-off $M$ and therefore it does not propagate. If $\gamma n^2 < 0$, the vector-boson mass-term has the wrong sign and the vacuum with $\langle A^{\mu} \rangle =0$ is unstable: the modes are tachyonic. In the Appendix, we show that for fluctuations about the vacuum $\langle A^{\mu}\rangle = M n^{\mu}$, there are three massless modes polarized orthogonally to $n^{\mu}$ in the long-wavelength limit. This fits with Goldstone's theorem: three of the six antisymmetric generators  ${\mathcal{J}}_{\mu\nu}$ of the global Lorentz group are broken by the order parameter $n^{\mu}$ and there should therefore be three massless long-wavelength excitations with polarization $\epsilon_{\mu}={\mathcal{J}}_{\mu\nu} n^{\nu}$, which  satisfy $\epsilon \cdot n = 0$  by antisymmetry of the Lorentz generators. In the case of spacelike $n^{\mu}$, there is also a mode with frame-dependent mass $O(gM)$. We note that the theory is perturbatively stable for both timelike and spacelike $n^{\mu}$, even though the potential is not bounded below in the timelike case. The same is true for the theory coupled to gravity, which we discuss in the next section.

\subsection{With gravity}
We now couple the vector field to gravity. Discarding any cosmological constant, there is a vacuum with a flat background and  non-zero \emph{vev} for the vector field. The fluctuations can be written as
\begin{align} 
A^{\mu} &= Mn^{\mu} + g B^{\mu}, \nonumber \\
g_{\mu\nu} &= \eta_{\mu \nu} + h_{\mu\nu}/{M_P}. \nonumber
\end{align}
\subsubsection{Gauge fixing}The linearized co-ordinate transformations are \cite{note5}
\begin{align} \label{gt}
x^{\mu} &\rightarrow x^{\mu} + \xi^{\mu}, \nonumber \\
h_{\mu\nu} &\rightarrow h_{\mu\nu} - \partial_{\mu}\xi_{\nu} - \partial_{\nu}\xi_{\mu},  \nonumber\\
gB^{\mu} &\rightarrow gB^{\mu} + M n^{\sigma} \partial_{\sigma} \xi^{\mu}.
\end{align}
A convenient gauge condition is the \emph{decoupling gauge} \cite{note6}
\begin{gather} 
B^{\mu} = 0. \nonumber
\end{gather}
 The residual gauge transformations which preserve this gauge satisfy 
\begin{gather} \label{rgt}
n^{\sigma} \partial_{\sigma} \xi^{\mu} = 0.
\end{gather}
The gauge-fixed Lagrangian for quadratic fluctuations in momentum space is, from (\ref{act})
\begin{widetext}
\begin{multline} \label{act2}
\mathcal{L} = \frac{1}{2} h_{\mu\nu} \Bigg[
-\frac{k^2}{2}\eta^{\mu\nu}\eta^{\sigma\rho}
+\frac{k^2}{2}\eta^{(\mu\sigma}\eta^{\nu\rho)}
+\eta^{(\mu\nu}k^{\sigma}k^{\rho)}
- \eta^{(\mu\sigma}k^{\nu}k^{\rho)}
-\frac{M^2}{2g^2{M_P}^{2}}\bigg(\eta^{(\mu\sigma}-k^{(\mu}k^{ \sigma}\bigg)n^{\nu}n^{\rho )} \\
-{\alpha_1} \frac{M^2}{{M_P}^{2}} \bigg( 
\frac{(n \cdot k)^2}{4}  \eta^{\mu\nu}\eta^{\sigma\rho}
-\frac{(n \cdot k)^2}{4}  \eta^{(\mu\sigma}\eta^{\nu\rho)}
-\frac{1}{2} n^{(\mu}n^{\sigma} k^{\nu}k^{\rho)}
+\frac{k^2}{2} n^{(\mu}n^{\sigma} \eta^{\nu\rho)}
\bigg) 
+\frac{{\alpha_2} (n \cdot k)^2}{4} \frac{M^2}{{M_P}^{2}} \eta^{\mu\nu}\eta^{\sigma\rho}\\
+\beta_2 \frac{M^2}{4 {M_P}^{2}} (n \cdot k)^2 \eta^{(\mu\nu} n^{\sigma}n^{\rho)} 
+\beta_3 \frac{M^2}{4 {M_P}^{2}} \bigg(+k^{2} n^{\mu}n^{\nu}n^{\sigma}n^{\rho} -2(n \cdot k)k^{(\mu}n^{\nu}n^{\sigma}n^{\rho)} 
 \bigg) 
+\beta_4 \frac{M^2}{4 {M_P}^{2}}(n \cdot k)^2n^{\mu}n^{\nu}n^{\sigma}n^{\rho}\\- \gamma \frac{M^4}{{M_P}^{2}} n^{\mu}n^{\nu}n^{\sigma}n^{\rho}
\Bigg] h_{\sigma \rho}.
\end{multline}
\end{widetext}
The parentheses in superscripts indicate symmetrization under the permutations $\mu \leftrightarrow \nu$, $\sigma \leftrightarrow \rho$ and $\mu \nu \leftrightarrow \sigma \rho$, 
rendering the kernel of the quadratic Lagrangian a symmetric linear operator on the ten-dimensional vector space of symmetric tensors with two indices.
\subsubsection{Graviton modes}
To find the modes of this modified gravity Lagrangian, one can (provided $n^{\mu}$ and $k^{\mu}$ are linearly-independent \cite{note10}) choose a basis of four-vectors $\{ k^{\mu}, l^{\mu}, m^{\mu},n^{\mu}\}$ such that  
\begin{gather}
k\cdot l = k \cdot m = l \cdot m = l\cdot n = m \cdot n =0. \nonumber
\end{gather}
From this basis of four-vectors, one can construct a basis for the ten-dimensional vector space of symmetric tensors with two indices, viz.
\begin{multline}
\{ k^{\mu}k^{\nu}, k^{(\mu}l^{\nu)} , k^{(\mu}m^{\nu)}, k^{(\mu}n^{\nu)},l^{\mu}l^{\nu}, \nonumber \\
 l^{(\mu}m^{\nu)},l^{(\mu}n^{\nu)},m^{\mu}m^{\nu},m^{(\mu}n^{\nu)},n^{\mu}n^{\nu}\}. \nonumber
\end{multline}
We now find the null eigenvectors of the kernel of the quadratic Lagrangian in (\ref{act2}). From the gauge transformations (\ref{gt}), it is clear that polarizations of the form $h_{\mu \nu} = k_{(\mu}\Xi_{\nu)}$ correspond to gauge transformations, where $\xi_{\mu} = \frac{i}{2} \Xi_{\mu} e^{ik\cdot x}$ and $\Xi_{\mu}$ is any constant four-vector. There are four such linearly-independent polarizations. Since we have fixed the gauge, these become null eigenvectors only when they correspond to residual gauge transformations, satisfying (\ref{rgt}). This is the case only if $n \cdot k = 0$. Explicitly computing the action of the kernel in (\ref{act2}) on these polarizations confirms that they are indeed null eigenvectors when $n \cdot k = 0$, as required. Moreover, these pure-gauge polarizations do not couple to conserved sources of the energy-momentum tensor, which satisfy $k_{\mu}T^{\mu \nu}=0$.

Two more eigenvectors of the Lagrangian kernel lie in the degenerate subspace spanned by
\begin{gather}\label{pol1} 
\frac{l^{\mu}l^{\nu}}{l^2} - \frac{m^{\mu}m^{\nu}}{m^2}\;\; \mathrm{and}\;\; l^{(\mu} m^{\nu)},
\end{gather} 
both with eigenvalue
$\frac{1}{2} \big( k^2 + \frac{{\alpha_1} M^2}{2{M_P}^{2}}(n\cdot k)^2\big)$. These are null eigenvectors when the dispersion relation
\begin{gather} \label{dis1}
k^2 + \frac{{\alpha_1} M^2}{2{M_P}^{2}}(n\cdot k)^2 = 0 
\end{gather}
is satisfied. These modes have the same polarizations as the two gravitons of GR \cite{Matsuki:1978rt}, but their dispersion relations are modified from $k^2 =0$ by corrections of $O(M^2/M_{P}^{2})$. They couple to sources of energy-momentum with standard strength $\frac{1}{M_P}$.

It is straightforward to show by explicit computation that two further null eigenvectors lie in the degenerate subspace spanned by polarizations of the form\begin{gather} \label{pol2}
\big(k_{(\mu} + \frac{\alpha_1 M^2}{2 M^{2}_{P}} (n \cdot k) n_{(\mu}\big) \epsilon_{\nu)},
\end{gather}
where $\epsilon^{\mu}=l^{\mu}$ or $m^{\mu}$ and the dispersion relation (to $O(M^2/{M_P}^2)$) is
\begin{gather} \label{dis2}
k^2 = 0.
\end{gather}
These two modes have no mass gap and consist of an $O(1)$ part, namely $k_{(\mu}\epsilon_{\nu)}$, which does not couple to conserved sources and an $O(M^2/M_{P}^{2})$ part, namely $n_{(\mu}\epsilon_{\nu)}$, which does. However, the coupling of this part to conserved sources is suppressed by a factor of $O(M^2/M_{P}^{2})$ relative to the coupling strength  $\frac{1}{M_P}$ of the other modes.

Any remaining null eigenvectors lie in the subspace spanned by 
$\{k^{\mu}k^{\nu}, k^{(\mu}n^{\nu)},n^{\mu}n^{\nu},\eta^{\mu \nu} \}$. Evaluating the action of the Lagrangian kernel on this subspace and setting the determinant of this linear transformation to zero, one recovers two of the pure gauge modes identified previously, and a further mode which is massive, but with a mass of the order of the UV cut-off $M$. This mode therefore does not propagate. 
\subsubsection{Positive-definiteness}
We now show that, for $\alpha_1 <0$, the four physical modes have positive-definite norms for all choices of the four-momentum $k^{\mu}$ and  the \emph{vev} $n^{\mu}$, such that the Hilbert space of physical states also has positive-definite norm.

The polarizations (\ref{pol1}) have norms two and $l^2 m^2$ respectively. Now $n^{\mu}$ cannot be null and $k^{\mu}$ is not null when (\ref{dis1}) holds, so the vectors $\{{k'}^{\mu}, l^{\mu},m^{\mu},n^{\mu}\}$ (where ${k'}^{\mu}=k^{\mu}-\frac{(n \cdot k)}{n^2}n^{\mu}$) form an \emph{orthogonal} basis. Since the Minkowski metric has signature $+---$, exactly one of these vectors has positive norm. From (\ref{dis1}), 
\begin{gather}
{k'}^2 = - \frac{(n \cdot k)^2}{n^2}\bigg(1+\frac{\alpha_1 M^2}{2 M_{P}^{2}}n^2 \bigg) \simeq - \frac{(n \cdot k)^2}{n^2}. \nonumber
\end{gather}
Thus, $n^2$ and  ${k'}^2$ have opposite sign and one of these must be the basis vector with positive norm. The other two basis vectors must have the same sign norm, such that $l^2m^2>0$. Both polarizations (\ref{pol1}) thus have positive-definite norm.

The remaining modes (\ref{pol2}) have norms given on-shell by $\alpha_1 \frac{M^2}{2{M_P}^2}(n \cdot k)^2 \epsilon^2$ to  $O(M^2/M_{P}^{2})$. Using $k^2=0$, arguments similar to those of the preceeding paragraph then indicate that $n^2$ and  ${k'}^2$ have opposite sign, such that the other elements in the orthogonal basis both have negative norm. Thus the modes (\ref{pol2}) both have the same-sign norm, which is furthermore positive if $\alpha_1<0$.
 
\section{Discussion}
In all, we see that there are four massless, physical graviton modes with positive-definite norm (for  $\alpha_1<0$) in this example. Only the parameter $\alpha_1$ results in physical deviations from GR in the weak field limit. Two of the modes become standard gravitons (with dispersion relation $k^2=0$) in the limit $M \rightarrow 0$ and the other two decouple from conserved energy-momentum sources in this limit. The absence of a discontinuity is therefore manifest and the physical predictions of the modified gravity theory in a region of spacetime which is approximately flat can be pushed arbitrarily close to the predictions of GR by choosing $M$ to be sufficiently small. Indeed, since the corrections are of order $M^2/M_{P}^{2}$, the upper bound for $M$ can be made very large, within a few orders of magnitude of the Planck scale, without observable deviations around flat backgrounds. This perhaps explains why broken Lorentz invariance in the gravitational sector has not yet been observed, despite the apparent `naturalness' of the effective field theories discussed here. A lower bound for $M$ can be found by requiring that the cut-off  lies beyond the reach of current short-distance tests of gravity, namely $O(10^{-2} \mathrm{cm})$. 

Finally, we re-iterate that even though the deviations of the theories presented here from GR are small in regions of spacetime which are approximately flat, there is no reason why the deviations should be small in regions of spacetime whose extent exceeds the curvature scale. The study of these theories in such regimes is a pressing question. Perhaps they may even play a role in the explanation of large-scale gravitational phenomena.
\begin{acknowledgments}
This work was supported by a PPARC grant. I thank N.~Arkani-Hamed, J.~Bjorken, S. Dubovsky, T.~Jacobson and especially J. March-Russell for discussions. I thank S. Dubovsky for pointing out a mistake in an earlier version.
\end{acknowledgments}
\appendix*
\section{The Goldstone Analogue}We expand
\begin{gather}
A^{\mu} = M n^{\mu} + g B^{\mu}, \nonumber
\end{gather}
such that (\ref{act}) becomes
\begin{widetext}
\begin{multline}
\mathcal{L} = \frac{1}{2} B^{\mu} \bigg[ - \eta_{\mu\nu}(k^2 - g^2(\beta_1 + \beta_3) (n \cdot k)^2) + k_{\mu}k_{\nu}(1+{\alpha_2} g^2) \\
+ k_{(\mu}n_{\nu)} g^2 (n \cdot k) (\beta_2 - 2\beta_1) 
+ n_{\mu}n_{\nu} g^2(-4\gamma  M^2 +   (n \cdot k)^2 \beta_4 + \beta_1 k^2) \bigg] B^{\nu}. \nonumber
\end{multline}
\end{widetext}
Two modes have polarization orthogonal to both $k_{\mu}$ and $n_{\mu}$, with Lorentz-variant dispersion relation
$k^2 - g^2(\beta_1 + \beta_3) (n \cdot k)^2 = 0.$
Since $\omega(\mathbf{0})=0$, they are massless in all frames.
The other two modes lie in the subspace spanned by $k^{\mu}$ and $n^{\mu}$. To
find their masses, set $\mathbf{k}=\mathbf{0}$ and set the determinant of the action of the kernel on the subspace equal to zero. At leading order in $g$ and for $\omega \ll M$, this relation reduces to
\begin{multline}
({\alpha_2} + \beta_3 {n^0}^2 + \frac{1}{2}\beta_2{n^0}^2)\omega^2 (\omega^2 + 4\gamma g^2 M^2 n^2) \\
 -4 (n^0)^2 \gamma \omega^2 M^2 =0, \nonumber
\end{multline}
implying $\omega^2 =0$ or $\omega^2 = -4\gamma g^2 M^2 n^2 + \frac{4(n^0)^2 \gamma M^2}{({\alpha_2} + \beta_3 {n^0}^2 + \frac{1}{2}\beta_2{n^0}^2)}$. The first solution is massless and is polarized perpendicularly to $n^{\mu}$ at $\omega = 0$. The second solution corresponds to a mode with frame-dependent mass. For spacelike $n^{\mu}$, such that $n^0 \simeq 0$, this mass is $O(gM)$ and the mode is physical; for timelike $n^{\mu}$, $n^0 \simeq 1$, and the mass is at the cut-off \cite{note8}.

\begin{thebibliography}{10}
\bibitem{note2}
The terms with coefficients $\beta_j$ are not suppressed in the vacuum where $\langle A^{\mu} \rangle = M n^{\mu}$; we have also neglected terms, such as $R A_{\mu}A^{\mu}$, which simply shift the coefficients of other terms for quadratic fluctuations about the vacuum.
\bibitem{note5}
Henceforth, all indices are understood to be raised and lowered with the flat metric $\eta_{\mu\nu}=\mathrm{diag}(1,-1,-1,-1)$.
\bibitem{note6}
To prove that this gauge is accessible from any field configuration ${B'}^{\mu}$, one has to show that
$0 = {B'}^{\mu} + \frac{M}{g} n^{\sigma} \partial_{\sigma} \xi^{\mu}$
admits a solution. It suffices to choose the four gauge functions $\xi^{\mu}$ such that their derivatives in the $n^{\sigma}$-direction cancel the four components of ${B'}^{\mu}$. Since the fluctuations of the vector field can be completely gauged away, there is no analogue of the Higgs boson in this example. This will not be true in general.
\bibitem{note10}
There are no physical modes in the case where $n^{\mu}$ and $k^{\mu}$ are linearly-dependent.
\bibitem{note8}
The components of $n^{\mu}$ could become arbitrarily large were we free to perform arbitrarily large Lorentz boosts. However, we have already imposed the observer-dependent cut-off $\omega \lesssim M$ and this restricts us to small boosts.
\end{thebibliography}

\end{document}